\documentstyle[prl,epsfig,aps]{revtex}
\begin{document}
\twocolumn[\hsize\textwidth\columnwidth\hsize\csname@twocolumnfalse\endcsname
\title{Aging in Models of Non-linear Diffusion}
\author{Daniel A. Stariolo}
\address{
Dipartimento di Fisica, Universit\`a di Roma I
{\em ``La Sapienza''}\\
 and\\
Istituto Nazionale di Fisica Nucleare, Sezione di Roma I\\
Piazzale Aldo Moro 2, 00185 Rome, Italy\\[0.3em]
\tt stariolo@chimera.roma1.infn.it
}
\maketitle

\begin{abstract}
We show that for a class of problems described by the non-linear diffusion
equation $\frac{\partial}{\partial t}\phi^{\mu}=D\,\frac{\partial^2}
{\partial x^2} \phi^{\nu}$ an 
exact calculation of the two time autocorrelation function gives 
 $C(t,t')= f(t-t')g(t')$, $(t > t')$ exhibiting
normal and anomalous diffusions, as well as aging effects, depending on the 
values of $\mu$ and $\nu$. We also discuss the form in which the Fluctuation-
Dissipation Theorem 
is violated in this type of systems. Finally, we argue that in this kind 
of models, aging may be consequence of the non conservation of the 
``total mass''. 

\end{abstract}

\twocolumn
\vskip.5pc]
\narrowtext

In a wide variety of physical systems where some kind of diffusion takes place
it can be observed that the mean-squared displacement scales with time as
$<x^2(t)> \, \propto \, t^{\alpha}$ with $\alpha$ depending on the physical 
problem in 
question. $\alpha=1$ corresponds to the so called {\it normal diffusion} (the
simple random walk), of
which a complete statistical description
can be obtained, for example, from the solution of the well known 
{\it diffusion equation}
$\frac{\partial}{\partial t}\phi(x,t) = D \frac{\partial^2}{\partial x^2}
\phi(x,t)$, where $\phi(x,t)$ is the probability that the diffusing particle
be at position $x$ at time $t$ provided it was at the origin $x=0$ at $t=0$ and
$D$ is the diffusion constant. If $\alpha \neq 1$ the diffusion is
called {\it anomalous} with $\alpha <1$ corresponding to {\it subdiffusion}
and $\alpha >1$ to {\it superdiffusion} \cite{bouchaud}. Anomalous diffusion 
can be consequence, for example, of some kind of disorder in the system 
\cite{bouchaud,zumofen}, or more generically, of longe range
correlations in space-time. The computation of the propagator $\phi(x,t)$, 
which contains all the
spatio-temporal information of the system, is in general a difficult 
task.
Without knowning
the exact propagator for all times, its long time form can be calculated 
in some cases using techniques like
renormalization group and scaling arguments \cite{goldenfeld1}.

Recently Tsallis {\it et.al.}\cite{tsallis1} have obtained the {\it exact}
solution of the non-linear Fokker-Planck equation
\begin{equation}
\frac{\partial}{\partial t}\phi(x,t)^{\mu}=
-\frac{\partial}{\partial x}\left\{F(x)[\phi(x,t)]^\mu \right\} +
D \frac{\partial^2}{\partial x^2}[\phi(x,t)]^{\nu}\,.
\label{general}
\end{equation}
where $(\mu,\nu) \in {\cal R}^2$, $D > 0$ is a diffusion constant, $F(x)=
-dV(x)/dx$ is an external force associated with the potential $V(x)$ and
$(x,t)$ is 1+1 space-time. They have found the solution for a drift of the
form $F(x)=k_1-k_2 x$ with $k_1$ and $k_2$ constants. This equation recovers
the standard diffusion or Fokker-Planck equation when $\mu=\nu=1$. 
Other values of $(\mu,
\nu)$ represent interesting physical systems as well: in the case with
$F(x)=0$ (a purely diffusive problem), for $\mu=1$ and arbitrary
$\nu$, Eq.(\ref{general}) is known as the {\it Porous Medium Equation} and
models many non-equilibrium systems in fluid dynamics \cite{aronson}, 
particle diffusion in magnetic fields \cite{berryman} and gas dynamics 
\cite{zeldovich}, depending on
the value of $\nu$. The information-theoretic aspects
of this case have been studied by Plastino and Plastino \cite{plastino} and
recently the interplay between dynamic and thermodynamic aspects have been
studied in \cite{compte} for the general case of diffusion in N dimensions.
The case $\mu=1$ and $\nu=3$ has
been studied by Spohn \cite{spohn} and describes a solid-on-solid model of
surface growth.

Restricting to the situation without drift (for the general case see
 Ref.\cite{tsallis1}),
the solution for the propagator $\phi_q(x,t)$ can be written as 
\begin{equation}
\phi_q(x,t)=
\frac{ \left[ 1-\beta(t)(1-q)[x-x_M(t)]^2 \right]^{\frac{1}{1-q}}}{Z_q(t)} \, ,
\label{prob}
\end{equation}
with $q=1+\mu-\nu$ and $x_M(t)=x_M(0)$ is the mean position, which for a
situation without drift is constant and equal to the initial position. 
This solution is closed by the relations satisfied by $\beta(t)$ and $Z_q(t)$,
namely:
\begin{equation}
\frac{\beta(t)}{\beta(0)}=\left[ \frac{Z_q(0)}{Z_q(t)} \right]^{2\mu}
\label{betazeta}
\end{equation}
and
\begin{equation}
Z_q(t)=\left\{ [Z_q(0)]^{\mu+\nu} + 
\frac{2\nu(\nu+\mu)D\beta(0)[Z_q(0)]^{2\mu}}{\mu}t \right\}^{\frac{1}
{\mu+\nu}}\,.
\label{zeta}
\end{equation}
A static form of Eq.(\ref{prob}) with $\beta(t)=1/T$ (inverse temperature) and
$Z_q(t)=Z_q(T)$ (partition function) has been obtained from a maximum entropy 
principle in the context of a generalized thermstatistics \cite{tsallis2}, and
succesfully applied for explaining, among many other problems,
 the thermodynamic foundations of L\'evy anomalous
diffusion \cite{andre,zanette}.
We will see in the following that the above solution presents 
a very rich dynamical behaviour characterized in general by anomalous 
diffusion and, for certain values of $\mu$ and $\nu$, by aging phenomena, the
long term memory effects observed and nowadays extensively studied in 
amorphous polymers
\cite{struik} and spin glasses \cite{vincent}. Let us consider the two times
autocorrelation function
\begin{eqnarray}
C(t,t') & \equiv\ & \langle y(t)x(t') \rangle \\
        & =       & \int_{-\infty}^{\infty} \, dxdy \,
x\,y\,\phi_q(x,0,t')\,\phi_q(y,x,t-t') \,,
\end{eqnarray}
in which $t' < t$ and
where $\phi_q(u,v,z-z')$ is the probability that the particle was at position 
$u$
at time $z$ provided it was at position $v$ at time $z'$. From Eq.
(\ref{prob}) we obtain
\begin{equation}
C(t,t') = K_q \left[ Z_q(t-t')Z_q(t')[\beta(t-t')]^{1/2}
[\beta(t')]^{3/2} \right]^{-1} \,.
\label{anterior}
\end{equation}
with $K_q$ a constant that only depends on $q$.
Now considering the regime in which $t-t' \rightarrow \infty$ and also 
$t,t' \rightarrow \infty$, from Eqs.(\ref{betazeta}), (\ref{zeta}) and
(\ref{anterior})
\begin{equation}
C(t,t') = A \left[B\,(t-t')\right]^{\frac{\mu-1}{\mu+\nu}} \,
\left[B\, t'\right]^{\frac{3\mu-1}{\mu+\nu}} \,.
\label{corr}
\end{equation}
where
\begin{equation}
A=\frac{\Gamma\left(\frac{1}{2}\right)\Gamma\left(\frac{1}{q-1}-\frac{1}{2}
\right)\Gamma\left(\frac{3}{2}\right)\Gamma\left(\frac{1}{q-1}-\frac{3}{2}
\right)}{\Gamma^2\left(\frac{1}{q-1}\right)(q-1)^2}
\frac{1}{\beta^2(0)}\frac{1}{Z^{4\mu}(0)}
\end{equation}
and
\begin{equation}
B=2\,D\,\frac{\nu(\nu+\mu)}{\mu}\,\beta(0)\,Z^{2\mu}(0)
\end{equation}

This result presents a variety of interesting characteristics. First 
we note that for $\mu=\nu=1$ we obtain the well known result for normal
diffusion $C(t,t') = 2\,D\,min(t,t')$. For the case of the
porous medium equation, {\it i.e.} $\mu=1$, the asymptotic correlation
simplifies to
\begin{equation}
C(t,t') = A \left[ B\, t' \right]^{\frac{2}{1+\nu}} \,.
\label{corrporo}
\end{equation}
In this case the long time behaviour depends only on the minimum time (as in
normal diffusion) but the diffusion is
anomalous with exponent $2/(1+\nu)$. When $\nu > 1$ the behaviour is {\it
subdiffusive} and for $\nu < 1$ it is {\it superdiffusive}. This qualitative
change can be conveniently observed in the shape of the propagator at a fixed time
$t^\star$ as shown
in Figure 1. In the subdiffusive regime, characteristic of the porous medium
equation, the propagator presents a ``wave front'' which expands with time as
\begin{equation}
|x_{wf}|=\frac{1}{\sqrt{\beta(0)(\nu-1)}} \frac{Z_q(t)}{Z_q(0)}
\end{equation}
with $Z_q(t)$ given by Eq.(\ref{zeta}). As $t\rightarrow \infty$ it flattens
as $|x_{wf}| \propto t^{\frac{1}{1+\nu}}$. The ``wavefront'' changes to
the exponential decay of
the normal gaussian diffusion when $\nu=1$ and then to superdiffusion 
characterized by power law tails in the propagator which decays for
$x \rightarrow \pm \infty$ and $t$ fixed as
\begin{equation}
\phi(x) \propto |x|^{\frac{2}{\nu-1}}\,\hspace{1cm} \nu<1 \,.
\end{equation}
Although the previous discussion considered $\mu=1$, it is valid in
the general case in which the different regimes are more conveniently
characterized by the parameter $q=1+\mu-\nu$ ($q<1,1$ and $>1$ corresponds to
subdiffusive, normal and superdiffusive behaviours respectively), and the 
relation  defines the possible choices for $\mu$ and $\nu$.

For $\mu \neq 1$ the autocorrelation function depends explicitly on 
both times, on all time scales, a feature characteristic of systems with
long term memory. These  effects of {\it aging} are common in disordered
systems (like, {\it e.g.} spin glasses) where
the time correlations present particular scaling forms (see \cite{vincent} and
refernces therein). Aging effects have 
also been studied in models without explicit disorder as scalar
fields, the XY model and spinodal decomposition \cite{let2} and also in 
random walks in disordered media \cite{enzo,bouchaud2}.
In order to compare
the present scalings with those found in spin glasses it is convenient to
normalize the correlation function in order to have $C(t,t)=1$. This 
normalization is natural in magnetic systems but it is not so for an
arbitrary random walk. In our problem a suitable definition may be
\begin{equation}
C_n(t,t')=\frac{C(t,t')}{\sqrt{C(t,t)C(t',t')}}\,.
\end{equation}
With this definition we obtain
\begin{equation}
C_n(t,t') \propto (t-t')^\frac{\mu-1}{\mu+\nu}\left(\frac{t'}{t}\right)^
\frac{3\mu-1}{2(\mu+\nu)}
\end{equation}
when $t,t'\rightarrow\infty$ and $t-t'\rightarrow\infty$. This normalized
correlation presents the commonly seen scaling with $t'/t$. The breaking of
time translation invariance can be clearly seen in the so called 
``displacement'' function
\begin{eqnarray}
B(t,t') & \equiv & \langle [y(t)-x(t')]^2\rangle \\
        & =      & C(t,t)+C(t',t')-2C(t,t') \,.
\end{eqnarray}
By definition $B(t,t)=0$. Writing $\tau=t-t'$:
\begin{equation}
B(t'+\tau,t')\propto t'^{\frac{3\mu-1}{\mu+\nu}}+(t'+\tau)^{\frac{3\mu-1}
{\mu+\nu}}
-2\tau^{\frac{\mu-1}{\mu+\nu}}t'^{\frac{3\mu-1}{\mu+\nu}}
\end{equation}
We see that in general $B(t'+\tau,t')$ depends on both $t'$ and $\tau$ and
consequently time translation invariance is broken. When $\mu=1$ and
$\tau \gg t'$
\begin{equation}
B(t'+\tau,t')\;\approx \;\tau^\frac{2}{1+\nu}\,.
\end{equation}
The displacement depends only on the time difference $\tau$ and time
translation invariance is recovered.

These results also reflect in the behaviour of a suitable generalization of the
Fluctuation-Dissipation Theorem (FDT) for systems that may never attain
equilibrium. This generalized form, firstly introduced for studying the
off-equilibrium dynamics of spin glasses \cite{leticia}, states that
\begin{equation}
R(t,t')=\beta \theta(t-t') X(t,t')\frac{\partial C(t,t')}{\partial t'}.
\label{fdt}
\end{equation}
where $R(t,t')$ is the response at time $t$ to an external force $h$ applied at
time $t'$
\begin{equation}
R(t,t')=\left. \frac{\delta\langle x(t)\rangle}{\delta h(t')} \right|_{h=0},
\label{resp}
\end{equation}
$\beta$ is the inverse temperature and $\theta(z)$ is the step function. The
function $X(t,t')$ measures the departure from FDT; if FDT is satisfied then
$X(t,t')=1$. In general FDT will be violated by a system that never reaches
equilibrium as in the models we are considering here, but it is nevertheless
instructive to analyze how the function
$X$ behaves. If we apply an external
perturbation $h$ at time $t'$, the form of the propagator remains the same as 
in Eq.(\ref{prob}), but now $x_M(t)$ satisfies the differential equation
\begin{equation}
\frac{d}{dt}x_M(t)=h\,\delta(t-t')\,,
\end{equation}
whose solution is $x_M(t)=x_M(0)+h\theta(t-t')$. According to Eq.(\ref{resp})
the response function is $R(t,t')=\theta(t-t')$. Now, for the case of the 
usual random walk ($\mu=\nu=1$), an explicit calculation from Eq.(\ref{corr})
gives for the correlation function $C(t,t')=2Dt'$. Noting that in the
standard Fokker-Planck Equation $\beta\rightarrow 1/D$, we obtain from
Eq.(\ref{fdt}) that $X(t,t')=1/2$. Consequently, for the simple random walk,
FDT is violated with a factor $X$ that is a constant \cite{let2}. 
In the case of the Porous
Medium Equation ($\mu=1$, arbitrary $\nu$), a similar analysis shows that
$X(t,t')=f(t')$, {\it i.e.}, the correction factor is a function of $t'$ only.
In the third case, for arbitrary $\mu$ and $\nu$ the function $X$ depends
explicitly on both times $t$ and $t'$. From this analysis it is clear that,
although all three cases violate FDT (as expected), the dynamics becomes more
complex as we go from the simple random walk, to the Porous Medium Equation and
finally to arbitrary ($\mu,\nu$), and this is reflected in the structure of the
function $X(t,t')$.

Finally and from other point of view, 
we argue that the absence (presence) of aging
phenomena is a direct consequence of the ``conservation of mass" 
(or not) in the
system. For our model it can be verified that \cite{tsallis1}
\begin{equation}
\int\,dx\, \phi_q(x,t) = \left[ \frac{Z_q(t)}{Z_q(0)} \right]^{\mu-1} 
\int\,dx\, \phi_q(x,0) \,,
\end{equation}
and consequently the norm of the distribution or ``total mass'' is independent
of time only if $\mu=1$. As suggested by Goldendfeld for the case of the normal
diffusion, the conservation of mass is directly connected with the fact that
asymptotically the system loses memory of its history and becomes unable to
distinguish between different initial conditions. The situation is completely
different if
the mass is not conserved, as for example, in the modified porous medium 
equation \cite{goldenfeld2} and in the problems we consider here when 
$\mu\neq1$. Although we are not able to
give a rigorous proof of this assertion, the different situations studied here
are in agreement with it.
Consequently it would be a direct connection between conservation of
the mass and aging in the class of systems considered: for our model equation
in the systems with $\mu=1$ the mass is conserved and the autocorrelation
function does not present aging behaviour. But if $\mu\neq 1$ there is no
conservation of the mass and aging effects are seen.

Concluding,  a number of interesting problems that can be modeled by
non-linear diffusion equations can be solved exactly and the diffusion presents
very different characteristics depending on the degree of non-linearity.
An exact 
calculation of the two time correlation functions shows that, besides the
expected anomalous diffusion, some systems may exhibit aging effects as found
in many other disordered systems. The aging satisfies particular scaling forms,
depending on the problem considered, which in principle could be compared with
experimental results. The different dynamical scenarios can be characterized
also studying the (violation of) Fluctuation-Dissipation Theorem.
These aging effects, in systems modeled by
partial differential equations, seem to be strongly related to the conservation
of the ``total mass''. It would be interesting to test the validity (or not) 
of this hypothesis in other models and also to study the extensions of these
results for dimensions higher than one.

\vspace{1cm} I would like to thank Constantino Tsallis, Leticia F
Cugliandolo and David S Dean for very useful discussions and suggestions.

\begin{figure}[htbp]
\begin{center}
\addvspace{1 cm}
\leavevmode
\epsfysize=250pt
\epsffile{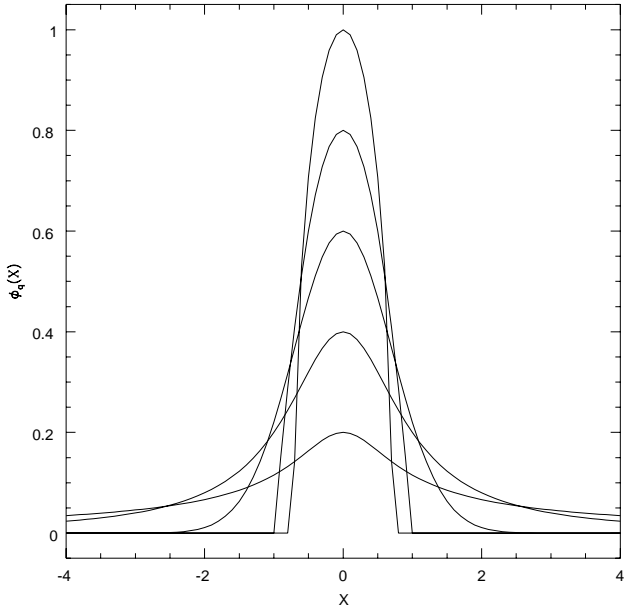}
\end{center}
\caption{Qualitative shape of the propagator $\phi_q(x,t)$ for fixed t. 
From top to bottom $q=-1,\,0,\,1,\,2,\,3$}
\label{fig}
\end{figure}


\begin{thebibliography}{15}
\bibitem{goldenfeld2} An interesting discussion of this and related questions 
can be found
in Ref.\cite{goldenfeld1} in the framework of the renormalization group.
\end{thebibliography}
\end{document}